# Development of soft and hard skills of high-school students via Young Physicists' Tournament

Sergej Faletič[1], Assen Kyuldjiev[2], Thomas Lindner[3,4], Dorottya Schnider[5], Éva Izsa[6], Mihály Hömöstrei[5], Péter Jenei[5], Boyka Aneva[2], Daniela D. Ivanova[6], Hynek Němec[7], František Kundracik[8], and Martin Plesch[9,10,*]

[1] University of Ljubljana Faculty of Mathematics and Physics, Jadranska 19, 1000 Ljubljana, Slovenia
[2] Institute of Nuclear Research and Nuclear Energy, Bulgarian Academy of Sciences, Tzarigradsko chaussee 72, Sofia 1784, Bulgaria
[3] Faculty of Business, University of Innsbruck, Karl-Rahner-Platz 3, 6020 Innsbruck, Austria
[4] Copenhagen Business School, Department of International Economics, Government and Business
[5] Department of Materials Physics, Eötvös Loránd University, Pázmány Péter sétány 1/A, H-1117 Budapest, Hungary
[6] Baba Tonka High School of Mathematics, Ivan Vazov 18, Ruse 7000, Bulgaria
[7] Institute of Physics of the Czech Academy of Sciences, Na Slovance 2, 182 00 Praha 8, Czech Republic
[8] Comenius University, Faculty of Mathematics, Physics and Informatics, Mlynská dolina F1, 84248 Bratislava, Slovakia
[9] Institute of Physics, Slovak Academy of Sciences, Dúbravská cesta 9, 845 11 Bratislava, Slovakia
[10] Faculty of Natural Sciences, Matej Bel University, Banská Bystrica, Slovakia
* Corresponding author: martin.plesch@savba.sk

The Young Physicists' Tournament (YPT) inspires high-school students to immerse themselves into an inquiry process closely resembling the real physics research. Using questionnaires' replies from the students and their teachers engaged in YPT, we investigated the perception of teachers and students on how the preparation for YPT contributes to the development of hard skills (physics content knowledge, mathematics, modelling...) and soft skills (communication, team work, organization...) of the students, and how it compares to regular classes and other activities. This comparison shows a positive role of YPT among students, and even a more positive perception among teachers. The significant development of a wide range of advanced skills justifies the substantial effort, resources and dedication of both teachers and students required for the participation in YPT.

*We condition students to expect simplicity. When they encounter complexity they may feel betrayed, disillusioned or "simply" lack the skills to interpret the circumstances [1]*

## I. INTRODUCTION

Young Physicists' Tournament (YPT) is a team competition for high school students, which closely mimics the process of authentic physics research [2 – 6]. It is also one of the very few international events which put emphasis on the students' inquiry activities. Despite the long history of the International Young Physicists' Tournament (IYPT, https://www.iypt.org) [7], there is little evidence on how the preparation for the Tournament influences students' development, especially of their hard and soft skills. To fill this gap, we performed a survey in five European countries (Bulgaria, Czechia, Hungary, Slovakia and Slovenia) among students and teachers engaged in YPT, asking them to assess the role of YPT, regular classes and other activities for the development of a selected set of hard and soft skills.

The following sections discuss YPT's features, educational relevance, and survey results. In section II we recall the unique features of the YPT as well as the skills required for the success in this competition. In section III we comment on the positioning of the YPT with respect to inquiry-based learning (IBL) and other modern educational concepts. Section IV is the key part of this paper devoted to the presentation of the principal outcomes of our survey and their discussion. The paper is summarized in Section V.

## II. YOUNG PHYSICISTS' TOURNAMENT (YPT)

This high-end competition was conceived in the specific intellectual atmosphere of Moscow extracurricular schools intended to look for and nurture talents in mathematics and other sciences. The engagement of top-level scientists naturally brought methods, research topics and ethos from research institutions. IYPT [7] as a multinational competition was born in 1988, and it quickly outgrew its origins to become a truly international endeavour. The initial idea to bring the "real-life" research experience to secondary school students has remained a symbol of YPT throughout the 35 years of its existence. IYPT was awarded the *ICPE medal* by the International Commission for Physics Education (ICPE), Commission 14 of the International Union of Pure and Applied Physics in 2013 for recognition of its "outstanding contributions to international physics education".

YPT combines a long-term team work on solving given open-ended problems with a debate about the obtained solutions, where each team rotates in the roles of Reporter (presenting its solution), Opponent (opposing the solution presented by the Reporter) and Reviewer (assessing the performance of the Reporter and the Opponent). The performance of the teams in each role is graded by a jury of physicists. The success in the competition is thus connected to a broad spectrum of hard and soft skills. The solution of the annual set of 17 problems involves several months of most scientists' real-life activities, including bibliographical research, designing experiments, data analysis, interpretation of measured data or development of own theoretical model. All results and conclusions must be ready before the graded debate. The YPT problems are always open-ended, therefore teams cannot reach any pre-defined "correct answer". Participants thus have to draw and support their conclusions argued from their experiments' outcomes. During the debate with peers, they have to defend the methodology used and the validity of their results. Critical assessment of results is thus another essential skill. Good English skills are also required, since this is the language of all performances in IYPT (and in most YPT national-level rounds, too); also, an important share of scientific literature in physics is available in English only. Finally, teamwork is required since YPT is a competition of quinary teams; nevertheless, an individual initiative of each team member is important, too.

The character of YPT and namely its contrast to popular competitions like *Physics Olympiads* oriented on solving close-ended problems is best illustrated by a YPT task assignment [8]. Here we state the problem №10 – *Water Rise* from the 26$^{th}$ IYPT:

*Fill a saucer up with water and place a candle vertically in the middle of the saucer. The candle is lit and then covered by a transparent beaker. Investigate and explain the further phenomenon.*

This is in fact the observation by Philo from Byzantium (3$^{rd}$ century BC) and it is the oldest reported phenomenon, by which a YPT task was inspired. The seeming simplicity of the phenomenon's demonstration is treacherous and the explanation which often comes first to mind (that the cause of the effect is the pressure decrease due to oxygen depletion) is actually wrong: a properly designed experiment is needed to resolve the true cause of the phenomenon [9]. Students not only have to propose explanations for the phenomenon, they also have to test them. One way to do this is to develop (or find in literature, if it exists) a physical and mathematical model that will allow them to calculate the change in the air pressure and therefore in water height and compare it to the actual experimental outcome. This engages students in the hypothetico-deductive reasoning, characteristic for scientific endeavours. If the predictions and outcomes do not match within the experimental uncertainties, students need to propose another explanation, or an explanation for the discrepancies. This requires a deep understanding of the model even if it was found in literature.

The YPT problems deal with phenomena which can be easily observed using the simplest equipment, but present an unexpected behaviour inexplicable through a trivial application of an established theory [6]. Unlike textbook exercises or typical lab works, they pique the curiosity of students and teachers alike. The tasks are also deliberately stated in a very broad way (e.g. "Investigate the phenomenon") which opens the door for different interpretations, allowing teams to select their own route to tackle the same problem [10]. The teams are mostly judged on the depth of their investigations and so students are often challenged to go far beyond the established theories to develop their explanations.

## III. YPT AND MODERN EDUCATIONAL CONCEPTS

The style of YPT problems invokes a natural association with *Inquiry Based Learning* (IBL) and *Inquiry Based Science Education* (IBSE) [11, 12]. It should be noticed that YPT was born long before most of these modern educational concepts became popular, and thus its development and maturing were self-guided, with no direct intent to apply specific didactic strategies. The broad scope of activities in YPT means that also other educational concepts like *Project-based Education*, *Problem-based Education*, *Active Learning* etc. may claim that YPT applies or even exemplifies their ideas. Among them, we see two that are particularly close to YPT: *Authentic Science* [13, 14], and *Investigative Science Learning Environment* (ISLE) [15, 16] which is not just a general idea but a well-proven learning system based on engaging students in a scientific reasoning process. Despite the recent surge of interest in all these IBL-related approaches, IYPT remains the only major recognized international competition evaluating high-school students' skills in solving inquiry problems in physics.

While working on the YPT problems, students have to become familiar with concepts, theories, models, and mathematics which go far beyond high school requirements. They need to acquire knowledge and an in-depth understanding of *selected topics*, which means that mastering the nowadays standard high-school curriculum is a prerequisite rather than the goal of the process. The participants usually develop a very strong feeling of ownership of the problems they are working on. Together with the competitive setting, this helps them harness their ambition and keeps them focused for an long period of time (preparation times are 3 to 6 months for the regional/national competitions and 8 or more months for IYPT) During the preparation, students fully aggregate both "hands-on" and "minds-on" approaches.

While YPT-activities are often linked to extracurricular activities and preparation for competitions, successful attempts have been made to integrate YPT elements into regular classes. We will discuss these attempts in the Discussion section. This survey gives some insight into which aspects of YPT might give the most gains in the limited time of regular classes and the proposed implementations will be discussed as well.

## IV. OUR SURVEY: RESULTS AND DISCUSSION

### A. Methods

Evaluations of the benefits of science competitions are rare; relevant studies are not popular since only a small fraction of students participates in the competitions. Investigations of contests like YPT is even more challenging as these are highly demanding competitions preventing a massive participation of both students and teachers. It is thus not surprising that an established approach to perform surveys on competitions is lacking and investigations are done mainly on a case-by-case basis. Despite these complications, it was for example found that science Olympiads in Slovakia were the most frequent motivation factor for students to pursue a scientific career (even slightly overcoming teachers' influence) [18]. Also, the names of about 75% of the former participants in IYPT (as well as in *International Physics Olympiad*) from 1998 to 2004 appeared in scientific databases, thus indicating that they had chosen a career in sciences [19].

We were eager to get a more elaborate output than mere assertions like "85% of teachers find YPT effective". Despite the temptation to study the YPT's unique strengths, we settled on the less ambitious task of comparing the perceived usefulness of YPT with that of *regular classes* and *other activities*. The usual form of the question was as follows:

*Please self-evaluate your competence/confidence in the given area in comparison with your classmates and indicate, which of the mentioned activities contributed the most to your development in this area.*

This formulation was chosen after careful consideration in collaboration with physics education experts trained also in assessing soft skills. Rubrics were considered as an assessment tool, but a major concern was that the problems of IYPT and other high-end competitions, such as the Physics Olympiad, are so diverse that it would be impossible to construct an objective descriptor of a skill level which could be used across various problems and various competitions. A problem-specific assessment is almost impossible since most respondents had experience with only one or a few problems.

For example, students may develop their "computer modelling" skill by numerically integrating a discretely measured function or by developing a visual simulation of the motion of a complex physical system, such as interacting fidget spinners. In both cases, their computer modelling skills would be developed by the activity, but it would be extremely difficult to provide a descriptor that would cover both types of modelling skills and anything in-between. The problems of YPT are just too different. That is why we opted for comparative assessment, asking students to rank their skills in comparison to their classmates' skills and to compare activities between themselves in terms of which activity contributed the most to their development of a particular skill.

We analysed these scores in terms of paired differences: the difference between the impact assigned to YPT activities and regular class activities, and the difference between the impact assigned to YPT activities and other extracurricular activities. The advantage of taking differences between the evaluations of the activities is that it avoids the need of using the absolute values of the grades, which may be highly compromised since different respondents may have different criteria and expectations. In this sense our investigations can serve as a pilot study to guide future surveys aimed at producing a sharper view of the YPT's strengths. The drawback is that the range of questions in the survey is limited to subjects simultaneously encountered in YPT, *regular classes* and in *other activities*.

In the present survey, we were asking teachers and students on how they perceived the contribution of YPT, *classes* and *other activities* to the development of selected hard and soft skills. The investigated skills can be divided into three categories:

1. *High-level hard skills*, presumably relevant for success in YPT:
- Designing experiments
- Interpreting experimental data, data analysis
- Developing own theoretical model
- Numerical simulations
- Independent research in scientific literature
- Critical assessment of others' results

2. *Curriculum-related hard skills* (lower-level skills usually associated with classical education):
- High school mathematics
- High school physics
- Solving close-ended problems in physics
- Conducting experiments based on a clear manual ("cookbook experiments")

3. *Soft skills*:
- Teamwork
- Ability to locate and use information
- Creativity
- Presentation skills

- Debating skills
- English (language) skills or ability

This list covers a reasonably wide range of skills (compared, e.g. with the list of skills in [2]) and allows us to pinpoint what are the strengths of YPT.

Student participants were selected based on their current or previous participation in YPT activities. Efforts were made to reach a reasonably large number of participants. For the teacher population we also reached those that were involved in YPT activities. As their number is naturally comparatively lower than the number of students, we decided to aim on personalized interviews. Although similar questionnaires were prepared for teachers and students in all countries participating in the survey, a few differences essential for the interpretation of the data had to be considered:

- Together with filling-in the survey, teachers were interviewed and therefore it was possible to guide them through the survey and explain unclear points. In particular, it was possible to clarify during the interview that the questions on regular classes meant in fact the regular physics classes, and that other activities analysed concerned other physics competitions. Students were filling questionnaires on the web instead; no further guidance was thus provided to them. Strictly speaking, it was thus possible that students considered not only regular physics classes, but other classes as well; analogically, under *other activities*, they could consider also activities not directly connected to physics. Nevertheless, from the context of the setting of the questionnaire, it is probable that they felt at least to some degree that activities in physics are in the centre of interest of the survey.
- Students rated the development of their skills on the scale –2 (no contribution of the activity to the skill development) to +2 (the activity contributes most to the skill developments). Teachers evaluated the development of student's skills on the scale 0 (= the activity does not develop the particular student's skill at all) to 10 (= the activity contributes a lot to the development of the particular student's skill).
- Teachers were interviewed in their mother language. Questionnaires for students were in their mother language, with the exception of Bulgarian students, who filled English questionnaires.

The differences between the students' and teachers' data collection is induced by their fundamentally different viewpoints on the subject. Students are asked what helped them, which they can know from personal experience, while teachers are asked what helped students, which they can only know from observation and conjecture.

On the other hand, students, in general, do not have expectations about how a particular activity should affect their learning, while teachers, by nature of their profession, should always consider the learning impact of students' activities.

The students' answers should, therefore, reflect their actual perception of what helped them the most. The teachers', on the other hand, might base their answers not only on the observed impact of the activity, but also on the expected impact and on the difference between them. To minimize this, all interviewers were instructed to pay attention that teachers' responses are based on observations rather than expectations.

**B. Results of the survey among teachers**

Practically all teachers in the participating countries having a first-hand acquaintance with the Tournament were invited for a phone-call or personal interview. Since a majority of them (32) accepted the invitation and completed the questionnaires, the survey had the potential to provide definitive results in this sample group. The survey took place in Bulgaria (9 respondents, autumn 2021), Czechia (4 respondents, autumn 2021), Hungary (5 respondents, 2020), Slovakia (11 respondents, 2020) and Slovenia (3 respondents, autumn 2021).

The results (represented by red bars in Fig. 1) are straightforward and completely in line with the expectations. Teachers assess that all *high-level hard skills* of students (*Designing experiments*, *Interpreting experimental data and data analysis*, *Developing own theoretical model*, *Numerical simulations*, *Independent research in scientific literature*, *Critical assessment of others' results*) are developed by YPT considerably more than by their regular physics classes or by other physics competitions. Remarkable here is also the magnitude of the difference, exceeding one third of the full scale almost in all cases. This only underlines that YPT is an activity which develops these skills much more than any other school-related activity. Exceptional is the difference in *Critical assessment of other's results* when comparing the contribution of YPT to that of other competitions (exceeding one half of the maximum scale): this means that YPT is practically the only extracurricular activity which develops this skill.

A similar general pattern is observed also in the assessment of all the investigated *soft skills*. On average, the domination of YPT over the regular physics classes is slightly less prominent, but still reaching a prominent level. YPT dominates other physics competitions especially in *Presentation skills*, *Debating skills*, *English skills* and *Teamwork*. This again indicates that there is a fundamental lack of other competitions developing these particular skills.

A richer behaviour is observed for the *curriculum-related hard skills* which are not expected to be developed significantly by YPT. Teachers assess that the only skill which is clearly better developed both by either regular classes or other physics competitions is *Solving close-ended problems in physics*. The contribution of YPT to the development of most other skills in this category is perceived comparably to that of regular classes and other physics competitions. This indicates an interesting unexpected result that YPT has its strong place even in the development of these competences usually associated with classical education methods.

The perception of the role of YPT on *Conducting experiments based on a clear manual* compared to the contribution of other physics competitions is surprising too. Here it is possible that teachers are aware that the amount of experimenting in most other physics competitions is so small that even the "routine" work within YPT significantly influences the development of students.

We reiterate that teachers' responses are based on their observations and conjecture. Therefore, they are based on perceptions and might be affected by expectations despite our efforts to minimize this effect. We will analyse how they relate to students' responses after we present both of them.

### C. Results of the survey among students

Targeted surveys were performed in autumn 2021. In total, 308 students from nine countries participated in the survey. However, this analysis was only done on a subset of 62 students who participated in a YPT competition or have declared solving at least one YPT problem.. The countries of origin included Bulgaria (6), Czechia (8), Hungary (7) and Slovenia (9). Included are also data from a similar survey performed in November 2020 during the preparatory seminar in Slovakia (32 respondents, all participating in some form of YPT competition; data on soft skills were not collected). Most of the students were in the last two years of high school (73%), with the rest in the third year from last (24%) or not answering the question (3%). Students had very different amounts of physics-based extracurricular activities, reporting between 0 and 11 hours per week with the median of one additional hour per week. 41% of students were female and 59% male.

A representative number of students varies by country, but, overall, we can cautiously assume that the participants in the analysis are representative of competitors in YPT competitions. The results for the entire ensemble of students are represented by blue bars in Fig. 1.

On the global level, students' comparison of YPT versus regular classes follows the same pattern as the comparison of teachers (i.e., a clearly significant contribution of YPT for the development of the *high-level hard skills* and *soft skills*, and a neutral assessment for the *curriculum-related hard skills*).

Somewhat lower enthusiasm for the YPT role may be at least partially attributed to the worse resolution in the scale of students' evaluation. Another reason may consist in the fact that—according to the phrasing of the survey—students could consider not only regular physics classes, but also other classes like specialized seminars. This latter view is strongly supported by the neutral assessment of the role of YPT for *English skills*: it is highly unlikely that English would be developed in physics classes, but it is strongly developed in regular English classes.

When comparing the contribution of YPT to that of other activities, there is a weak but clear influence of YPT in the case of *high-level hard skills*. The perception of YPT for the development of other *curriculum-related hard skills* as well as *soft skills* is within statistical errors comparable to that of other activities. Even this neutral results in fact emphasizes the remarkable impact of YPT, as its contribution is practically comparable to a broad spectrum of other activities possibly directly targeting some of the selected skills.

For a more detailed understanding of the students' responses, one could look into the results on a per-country basis (Fig. 2). We observe that the students are generally perceiving the role of YPT highly positively; only the Bulgarian students evaluate the role of YPT rather neutrally. This can be understood in different ways, ranging from the fact they did use an English version of the questionnaire to very small number of participating students (6). On the other hand, Czech students perceived most of the activities much above the average – this might be explained by the fact that a group of students with good knowledge and experience in YPT took part. As a curiosity one might point out that these students, unlike anyone else, consider YPT activities as very beneficial for cookbook experiments as well. We discuss this result further in the discussion.

All in all, while one has to be very careful in drawing conclusions for specific countries mainly due to small number of participating students, the students' evaluations integrated over all the countries provide valuable and significant results.

### D. Discussion

The different perceptions of teachers and students which we observe seem compatible with the view that students react to what takes place immediately in front of them and compare the gains with the morsels of knowledge they obtain in every lesson. Teachers, on the other hand, seem to pay more attention to long-term developments toward more abstract pedagogical notions. This is in line with the already noted fact that teachers and students tend to evaluate educational practices differently – the former do it mainly from an ostensive viewpoint while the latter from a performative one [20,21].

The most interesting results from the questionnaires are for the skills *High-school mathematics*, *Conducting experiments based on a clear manual*, *Teamwork*, *Ability to locate and use information*, and *Creativity*.

The contribution of YPT to *High-school mathematics* was assessed positively by the teachers, but negatively by the students. One possible explanation consists in the frequent use of calculus during the preparation for YPT, which is (at least by the older generation of teachers) considered to be a standard part of the high-school curriculum, while nowadays, it is taught superficially, if at all, and, therefore, not considered high-school mathematics by students.

Another possible explanation is that students underestimate the amount of high-school mathematics involved in YPT activities. Being usually good students, they may find these tasks trivial and thus not contributing to their development. Teachers, on the other hand, might recognize this amount but lacking first-hand experience of how much gain it brings to the students, overestimate its role in the development of the skill. Tasks that are easy do not contribute much to further development.

The contribution of YPT to *Conducting experiments based on a clear manual* was assessed mostly positively by teachers and students, which was a surprise to us, given the open-ended nature of YPT problems. We assume that we underestimated the role that procedures have in YPT problems. Complex and detailed procedures are often part of the YPT activities. To obtain reliable results for comparison with a model, students must often perform very specific procedures taking good care to vary only one parameter and to keep other conditions constant. The fact that students devise these procedures themselves does not change the fact that they are very detailed, manual-like procedures. Moreover, other tasks associated with cookbook labs are presentation of data and determination of uncertainties, both very important parts of YPT problems.

We expected that the contribution of YPT to *Teamwork*, *Ability to locate and use information*, and *Creativity* would be positive compared to regular classes and other activities, as was also the assessment of teachers. However, students assessed it neutral to negative when compared to their other activities. We believe that the explanation for this lies in the fact that *other activities* were not specified. Students might be engaged in activities like team-based sports, musical ensembles, debate club, arts, engineering project and similar activities, while teachers might have only thought of physics-oriented activities. Among the listed activities some put greater importance on teamwork and some greater importance on creativity than YPT.

As for the ability to locate and use information, the negative score of YPT against other activities is surprising. It is difficult to imagine an activity that would require more advanced search for information than YPT. One that comes to mind is the debate club, but we do not think that it can account for the difference. However, in the digital age, students might have compared YPT to any topic that they are passionate about and spend a lot of time researching on the internet. Clearly, this kind of information is much more multifaceted and from many more different sources than YPT information, so it could account for the difference. Yet, what we learn from this is that YPT is comparable in the development of these skills to activities for which these skills might be of crucial importance. This is an impressive result for YPT.

The positive role of YPT on the development of a broad spectrum of skills may be possibly utilized also in lower degrees in the undergraduate education. An especially positive effect was attributed to YPT in the skills of presentation and debating, so these skills will receive special focus.

The *YPT Toolkit* [17] collected examples of classroom practices to support teachers in integrating YPT-inspired activities also into their regular physics classes allowing more students to profit from this approach. It reports two implementations that use simpler physics experiments to create an environment of investigation. Both end with report and a peer opposition that attempts to gauge the strengths and weaknesses of the investigation. The above skills are strongly present in both implementations, along with the entire research process. A similar approach is implemented at the University of Ljubljana, where undergraduate physics students are offered a one-semester project laboratory course where they work on well-defined project tasks primarily based on YPT problems [22].

The scientific investigation part of the YPT model can and is being used in regular educational approaches. ISLE is built around mimicking the activities of actual scientists but with much simpler problems, such as cart collisions or scale balance, similar to the proposals in the YPT Toolkit. The crucial steps of observing, finding patterns, proposing explanations, making mathematical models, testing the models by comparing their predictions to the actual outcomes and providing explanations for any discrepancies are at the core of the ISLE approach. However, ISLE lacks the debate part. Given the simple nature of the problems, there is rarely disagreement among students or multiple solutions.

The implementation of the discussion element has been tried out in an engineering-oriented first-year university course with over 60 enrolled students, where groups of students propose the design of a device with given specifications and then present and discuss the proposals in a format mimicking the YPT discussion. The problems are, of course, less physically complex, but this is intentional to allow the students to understand the physics and challenge its implementation in the limited amount of time. The activity is completed in three hours with approximately five groups, with each group reporting and all other groups acting as opposition.

All implementations above sacrifice physical complexity and therefore the development of high-level hard skills for more development of soft skills. It is reasonable to expect that by passing from complex physics to more elementary physics, the development would shift more towards low-level hard skills, as the investigation process is still present in all of the implementations.

### E. Relation to the PISA conceptual framework

Somewhat unexpectedly, the mass-scale internationally standardised PISA educational assessment initiated by OECD became a major influencer on the worldwide education scene. Without promoting specific educational strategies, PISA began to mould the reference frame used to perceive modern learning and the competencies required. The PISA 2015, definition of the *scientific competencies* implies the abilities to [23]:
- *explain phenomena scientifically* – recognise, offer and evaluate explanations for a range of natural and technological phenomena;
- *evaluate and design scientific enquiry* – describe and appraise scientific investigations and propose ways of addressing questions scientifically;

- *interpret data and evidence scientifically* – analyse and evaluate data, claims and arguments in a variety of representations and draw appropriate scientific conclusions.

Practically all skills investigated here fall into these categories. From our results it is obvious that YPT is viewed by teachers as the activity which develops these competencies much more than regular physics classes or other physics competitions.

## V. CONCLUSIONS

The takeaway message of this paper is summarized in Fig. 3. Both students and teachers perceive that YPT develops the *high-level hard skills* and *soft skills* of the students fundamentally more than regular classes. On a scale where 4 is the maximum difference (between -2 and +2), and +2 is the maximum difference from neutral (0 to +2), the teachers scored this development at $1.6 \pm 0.1$ and $1.4 \pm 0.1$, respectively. Students scored this development a little lower at $0.9 \pm 0.1$ and $0.7 \pm 0.1$, respectively, so at about half the score of the teachers.

Teachers ranked similarly high the role of YPT in developing these skills also compared to other activities ($1.5 \pm 0.2$ for both), while students did not ($0.4 \pm 0.1$ and $0.0 \pm 0.1$, respectively). The most likely reason is that the teachers, being primarily engaged in physics, considered mostly other physics competitions, while students did most likely consider also activities outside physics. In light of this, it is impressive that they value the contribution of YPT activities as entirely comparable to activities which might be much more soft-skills oriented (team sports, debate club…)..

Assessment of the role of YPT for mastering the high-school curriculum is generally neutral ($-0.1 \pm 0.1$); the only skill where teachers and students alike do not see any significant role of YPT is the *solving close-ended problems*. This is an important finding because it shows that YPT does not compromise curriculum goals and the development of most of the low-level hard skills, while at the same time develops other skills much more than regular classes. Therefore, whatever the circumstances, YPT offers an inspiration for sensible inquiry activities suitable for various educational needs and desired outcomes.

The YPT-positive results of our survey may inspire further teachers to take up the YPT challenge. While it is clear that due to the nature of the problems and the tournament, not every student can participate fully in this high-end competition, many examples have been given of how it is possible to implement at least some of the important elements, such as scientific investigation, presentation and debate, in regular classes (as described e.g. in the *YPT Toolkit* [17] and in the examples above). While these implementations might shift the development from high-level hard skills towards low-level hard skills, this is exactly what a large class might most benefit from: the development of curriculum-based hard skills as well as soft skills among a larger mass of students.

## ACKNOWLEDGEMENTS

The present study was supported by the Erasmus+ project Dibali 2019-1-SK01-KA201-060798. MP acknowledges the support of projects 09I03-03-V04 00425 and 09I03-03-V04-00685 of the Research and Innovation Authority (Funded by the Next Generation EU). The work of HN was also supported by Operational Program Research, Development and Education financed by European Structural and Investment Funds and the Czech Ministry of Education, Youth and Sports (Project SOLID21 – CZ.02.1.01/0.0/0.0/16 019/0000760).

## ETHICAL STATEMENT

The authors acknowledge that all respondents participated voluntarily in the survey, and that a consent was obtained from all respondents to participate in the study.

## REFERENCES

[1] D. Allchin: How School Science Lies. In: *5th Annual Conference of the International History Philosophy and Science Teaching Group*, Como, Italy, p. 244 (1999)

[2] G. Tibell, Students' Skills Developed by Participation in the IYPT in: *What Physics Should We Teach? Proc International Physics Education Conference*, University of South Africa Press, p. 257 (2005).

[3] M. Plesch, F. Eller, C. Kanitz, J. Landgraf, A. Raab, and S. Selbach, Eur. J. Phys. **38**, 034001 (2017).

[4] M. Plesch, M. Badin, and M. Marek, Nature Sci. Rep. **8**, 3718 (2018); M. Plesch, M. Badin, and N. Ružičková, Eur. J. Phys. **39** 064003 (2018); M. Plesch, S. Plesník, N. Ružičková, Eur. J. Phys. **41** 034001 (2020).

[5] A. Kyuldjiev, IYPT Magazine **5**, 4 (2017).

[6] M. Tan, T. Koh, Science Education 1 (2022).

[7] The abbreviation IYPT denotes the official event of the International Young Physicists' Tournament, which takes place every summer with participating teams from about 30 countries. The abbreviation YPT denotes any regional, national or multi-national competition (like Austrian YPT) or training conducted in the IYPT manner and inheriting its problems.

[8] https://iypt.org/problems

[9] F. Vera, R. Rivera, and C. Nunez, Sci & Edu **20**, 881 (2011).

[10] https://archive.iypt.org/solutions

[11] A. Allinson, IYPT Magazine **1**, 4 (2013).

[12] We are aware that IBL and Inquiry Based Science Education (IBSE) are overused notions, which live together with a host of similar concepts, sometimes without clear boundaries between them. We shall use them mostly as placeholders for the numerous practices aiming to introduce inquiry (in one form or another) to school classes.

[13] B. Crawford: Authentic Science. In: *Encyclopedia of Science Education*, Springer, p. 113 (2015).

[14] B. Woolnough, School Science-Real Science? Personal Knowledge, Authentic Science and Student Research Projects. In: *Research in Science Education in Europe*, Springer, p. 245 (1999).

[15] E. Etkina, D. Brookes, and G. Planinsic, *Investigative Science Learning Environment*, Morgan & Claypool Publ, San Rafael, USA (2019).

[16] E. Etkina, G. Planinsic, and A. van Heuvelen, *College Physics: Explore and Apply* 2nd ed, Pearson, New York (2019).

[17] *YPT Toolkit*, http://dibali.sav.sk/wp-content/uploads/2020/07/IO1-document-toolkit_final_.pdf

[18] K. Hudecová, N. Pišútová, and J. Pišút, Československý časopis pro fyziku **62**, 472 (2012).

[19] M. Urbašíková: The impact of a development of ability to science process skills on choice of career in science. In: *DIDFYZ 2014 - Vymedzenie obsahu školskej fyziky,* UKF, Nitra, p. 337 (2014).

[20] T. Mostafa, A. Echazarra, and H. Guillou: The science of teaching science: An exploration of science teaching practices. In: *PISA 2015 OECD Education Working Paper № 188* (2018).

[21] M. Kloser, J. Research Sci. Teaching **51**, 1185 (2014).

[22] G. Planinsic, Eur. J. Phys. **30**, S133 (2009).

[23] D. Roberts and R. Bybee: Scientific Literacy, Science Literacy, and Science Education. In: *Handbook of Research on Science Education Vol II,* Routledge, New York, p. 545 (2014).

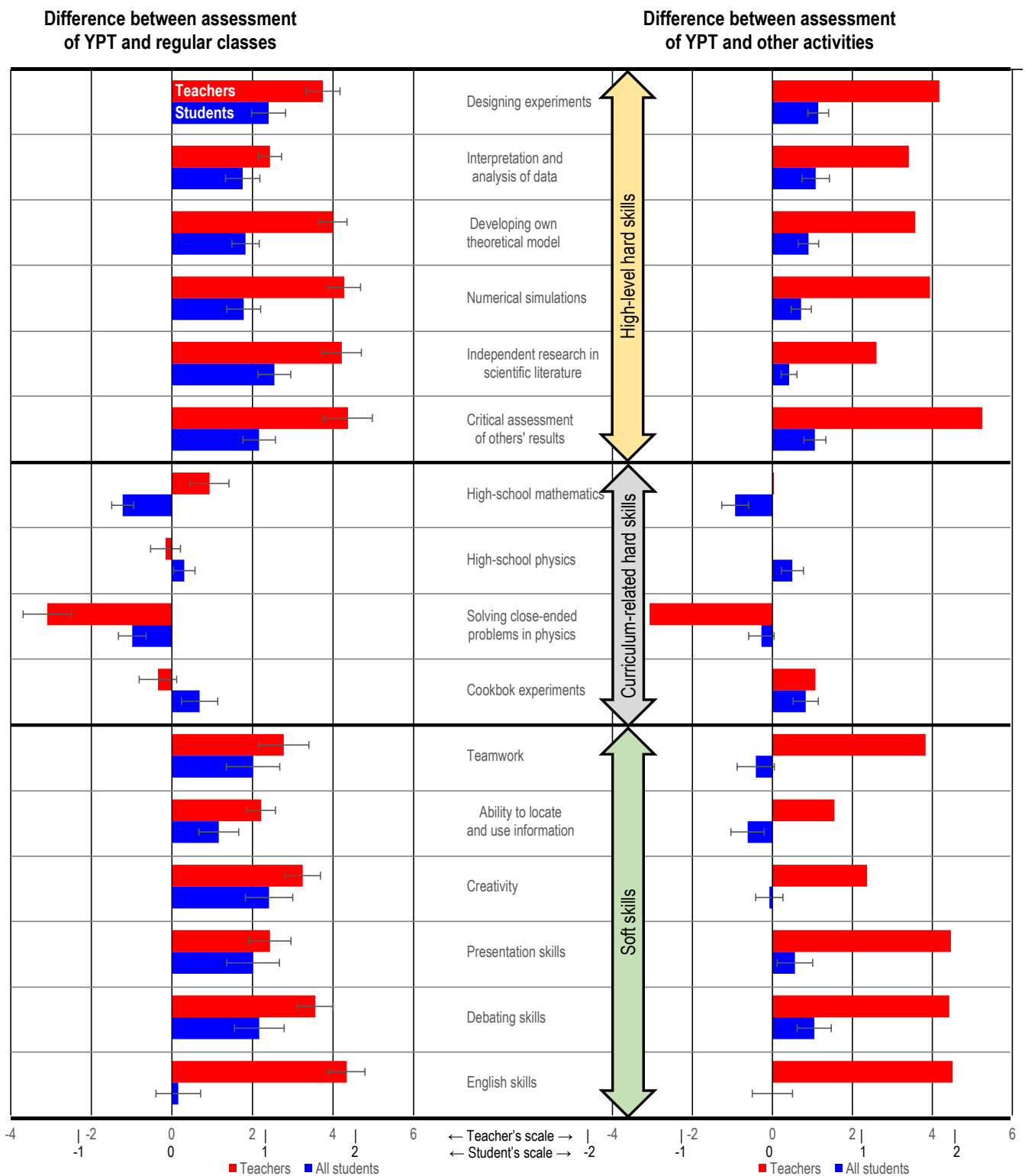

FIG. 1. Summary of the difference between the assessment of YPT and regular classes/other activities by teachers and students averaged over all countries. The teacher's and student's scale of responses were normalized to each other so that the maximum difference is represented by the same bar length in both sets of responses. Positive values indicate that the respondents assess that YPT contributes to the development of the given skill more than regular classes/other activities. The error bars represent the standard deviation of the mean value.

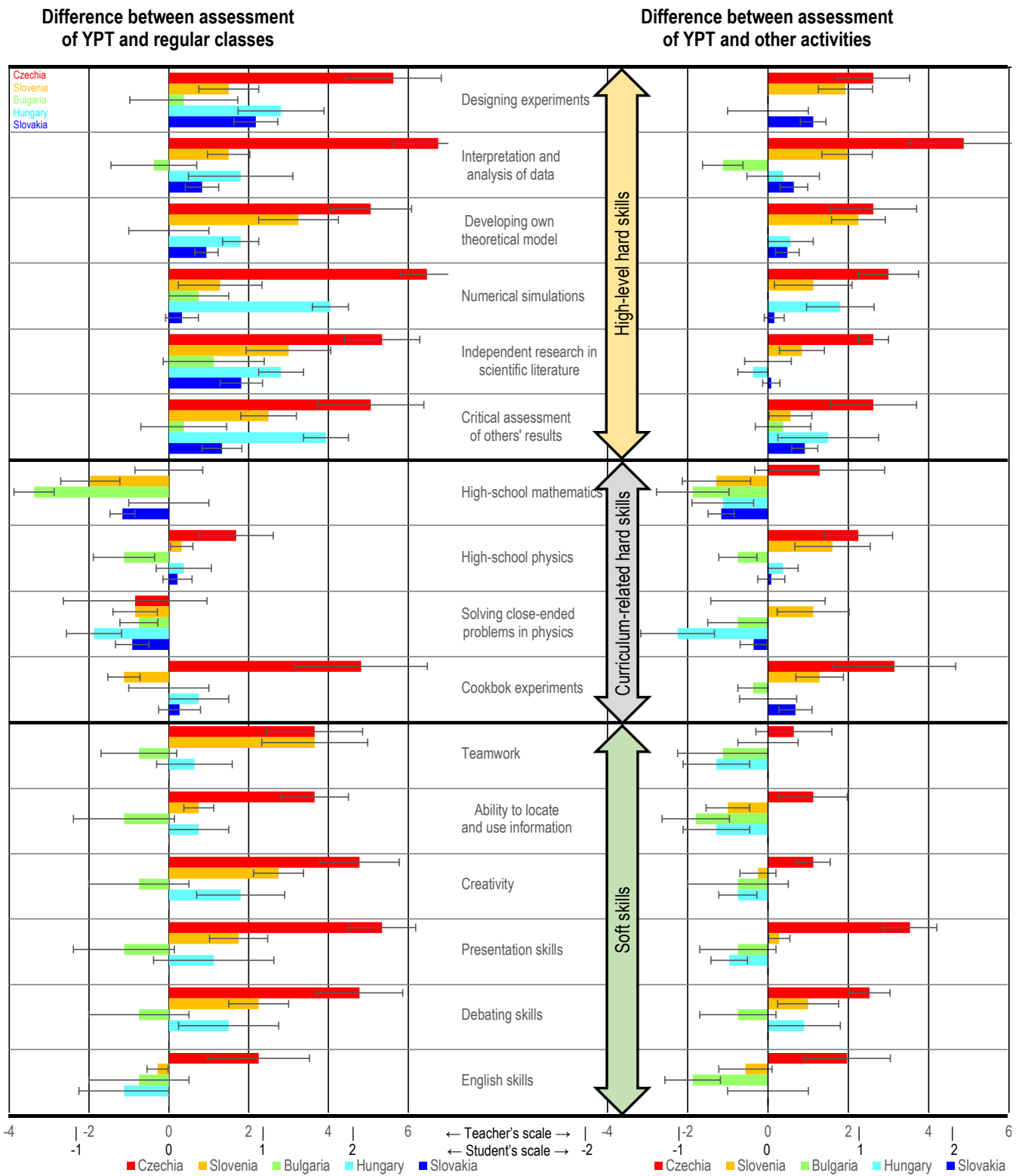

FIG. 2. Summary of the difference between the assessment of YPT and regular classes/other activities by students from various countries. The teacher's and student's scale of responses were normalized to each other so that the maximum difference is represented by the same bar length in both sets of responses. Positive values indicate that the respondents assess that YPT contributes to the development of the given skill more than regular classes/other activities. The error bars represent the standard deviation of the mean value.

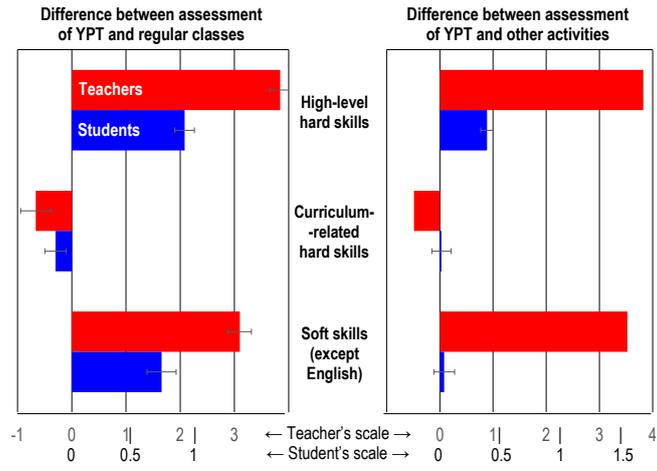

FIG. 3. Summary of the difference between the assessment of YPT and regular classes/other activities by teachers and students aggregated into the three major categories of skills.